\documentclass[12pt]{iopart}

\usepackage[]{units}
\usepackage{acronym}
\usepackage[T1]{fontenc}
\usepackage{lettrine} % The lettrine is the first enlarged letter at the beginning of the text
\usepackage{gensymb} 
\usepackage{float} 
\usepackage{nomencl}
\usepackage{caption}
\usepackage{subcaption}
\usepackage{ragged2e}
\usepackage{multirow}
\newcommand{\ssymbol}[1]{^{\@fnsymbol{#1}}}
\makeatother
\usepackage{graphicx}
\usepackage[colorinlistoftodos]{todonotes}
\usepackage{hyperref} % For hyperlinks in the PDF
\usepackage{booktabs,caption}
\usepackage[flushleft]{threeparttable}
\hypersetup{colorlinks = true,linkcolor = blue,anchorcolor =red,citecolor = blue,filecolor = red,urlcolor = blue}
\graphicspath{ {./img/} }
\acrodef{IR}[IR]{Infra-Red}
\acrodef{INFN}[INFN]{Istituto Nazionale di Fisica Nucleare}
\acrodef{PCTO}[PCTO]{Percorsi per le Competenze Trasversali e per l'Orientamento}
\acrodef{PLS}[PLS]{Piano Lauree Scientifiche}
\acrodef{RC}[RC]{Resistive-Capacitive}
\begin{document}

\title[Measurement of the Newton's cooling law time-constant by Arduino: an idea for STEM education in High Schools]{Measurement of the Newton's cooling law time-constant by Arduino: an idea for STEM education in High Schools}

\author{Fausto Casaburo}

\address{Sapienza Università di Roma, Dipartimento di Fisica}
\address{Istituto Nazionale di Fisica Nucleare (INFN) Sezione Roma}
\ead{fausto.casaburo@uniroma1.it-fausto.casaburo@roma1.infn.it}
\vspace{10pt}
\begin{indented}
\item[]March 2022
\end{indented}

\begin{abstract}
Commercial electronic measuring devices used in physics usually have high costs. In recent years, thanks to its low cost and the high number of available sensors, the Arduino board has been used for many educational purposes in physics education. In this paper a method to measure the Newton's cooling law time-constant by the Arduino board is presented.

\end{abstract}

%Uncomment for keywords
\vspace{2pc}

 %Uncomment for Submitted to journal title message
%\maketitle

%\linenumbers

%\tableofcontents

\section*{Introduction}\label{Introduction}
\addcontentsline{toc}{section}{Introduction}
\lettrine[nindent=0em,lines=3]{D}espite laboratory practice being essential for physics, due to lack of time and/or resources, it is often  neglected at High School. Many schools do not have adequately equipped laboratories.

During the last few years, many science communicators published educational articles that can easily made by students of High Schools using the Arduino \cite{Arduino} board. Some examples of these articles regard a body in free-fall \cite{casaburo2021teaching}, the  time constant of a \ac{RC} circuit \cite{Pereira_2016}, the kinetic friction coefficient \cite{_oban_2020}, and many more. 

Arduino is an open source platform made of electronic boards, sensors and expansion boards.  With an Arduino, students can acquire additional competences, as for example coding and programming \cite{Organtini_2018, Organtini_fisica_arduino}. Original Arduino boards can be bought very cheaply, but there are even cheaper clones available. Both the board and the sensors can be easily bought on-line or in electronics stores \cite{Organtini_fisica_arduino}. There are also many kits including the board and common sensors available for just 50-60 euro. Thanks to the low cost, the Arduino board and related components can be bought directly by students or by schools to be provided to students.

In this paper we present an Arduino-based experiment on Newton’s cooling law, allowing High School students to replicate it.

It consists of the measurement of the cooling time-constant of a liquid. By doing the experiment, students have the opportunity to show the exponential trend of the cooling law, estimate uncertainties, interpolate data and work with the exponential function. The latter is is an important function for physics also describing other physics laws as in the \ac{RC} circuit.

\section{Theory}\label{Theory}
Newton’s law of cooling (Eq. \ref{eq:Newtons_cooling}) describes the rate at which an exposed body changes temperature by radiation. According to this law, the trends over time of the temperature depends on the initial temperature of the body $T_0$ and the environment temperature $T_{env}$  as:
\begin{equation}
T\left(t\right)=T_{env}+\left(T_{0}-T_{env}\right)e^{-\nicefrac{t}{\tau}}
\label{eq:Newtons_cooling}
\end{equation}

Defining $m$ and $c$ as respectively the mass and the specific heat of the body, the time-constant is given by \cite{Vollmer_2009, DAVIDZON20125397}:
\begin{equation}
\tau\propto mc
\label{eq:tau}
\end{equation}

The value of the time constant depends on the composition of the body (and its container) and represents the time after which the the difference  $T_0-T_{env}$ is reduced to approximately  63\% of its original value. After $5\tau$ we expect that the body reaches room temperature.  It is similar to an \ac{RC} circuit. The exponential dependency of the temperature on the time can be verified and the time-constant $\tau$ can be estimated measuring the temperatures of body and environment. 

\section{Experimental setup and procedure}\label{Setup}
\begin{figure}[H]
\centering
\includegraphics[width=4in]{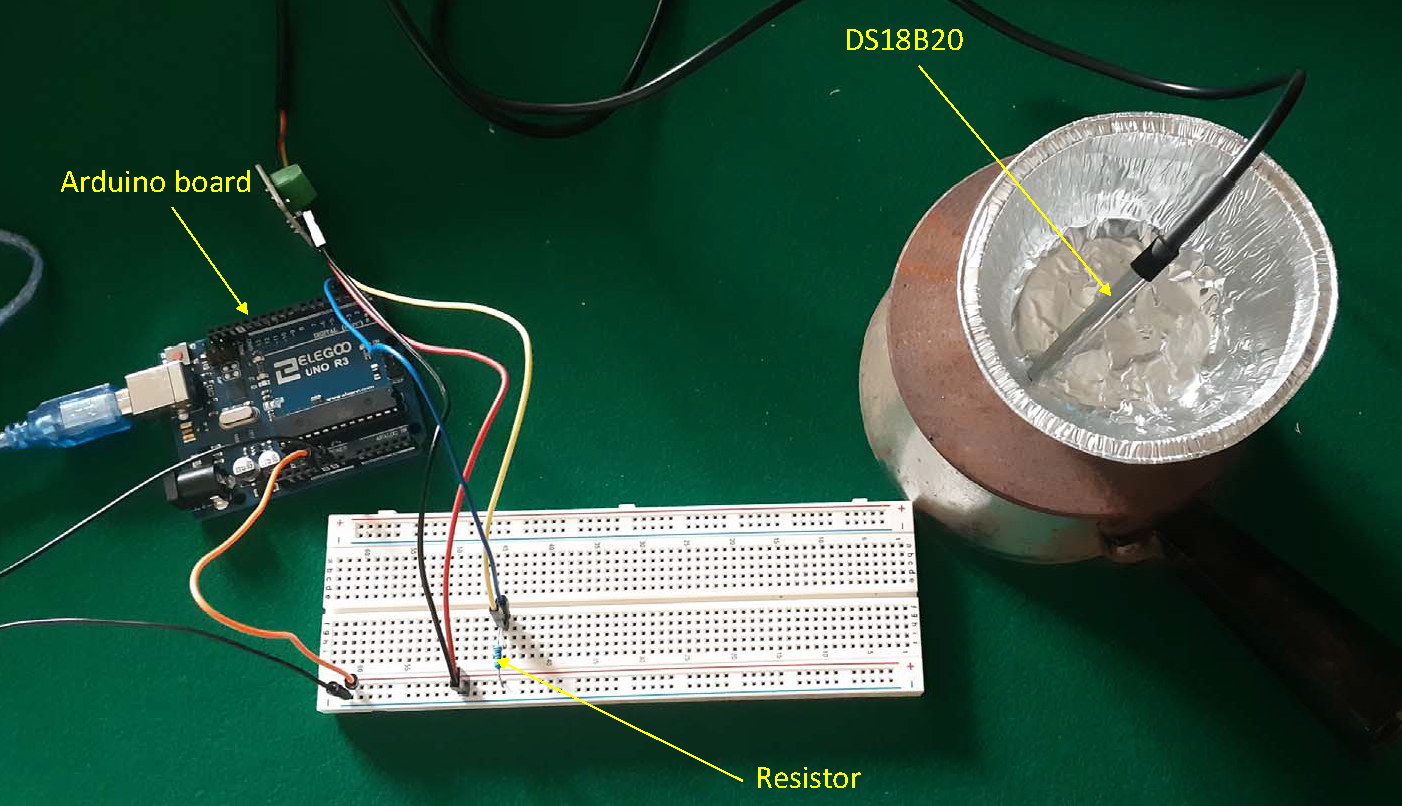}\DeclareGraphicsExtensions.
\caption{Experimental setup. }
\label{fig:exp_setup}
\end{figure}

The experimental setup (Fig. \ref{fig:exp_setup}) consists of an Arduino UNO R3 board, a temperature sensor DS18B20, a $\unit[10]{k\Omega}$ resistor, a room thermometer, a  stove, a computer, a breadboard, Dupont cables and a cup. The working range of the DS18B20 is $\left[\unit[218.15]{K};\unit[398.15]{K}\right]$ and the sensitivity is $\pm\unit[0.5]{K}$ in the range $\left[\unit[263.15]{K};\unit[358.15]{K}\right]$.
The resistor, the breadboard and the Dupont cables could be found in a typical Arduino kit. The sensor temperature must be purchased separately.

The temperature sensor is connected to the Arduino by Dupont cables. The sensor has three cables: the first one to be connected to the GND pin of the  Arduino, the second one to the $V_{cc}$ pin (both the $\unit[3.3]{V}$ or $\unit[5]{V}$ can be used) and the last one to be connected to a digital one. The resistor is inserted between the digital and the $V_{cc}$ pins as shown in Figure \ref{fig:exp_setup}. Lastly, the Arduino board is connected to the computer for data acquisition by the USB cable. The cup has been filled with a mass $m=\unit[\left(9.9\pm0.1\right)\cdot10^{-2}]{kg}$ (the specific heat of water is $c=\unit[4186]{\frac{J}{kg\cdot K}}$) and has been placed on the stove to be heated. The temperature sensor has been inserted in the cup to measure the water temperature during the cooling.

The water has been heated up to $T_{0}=\unit[\left(349.0\pm0.5\right)]{K}$, then it has been left cooling up to $T_{f}=\unit[\left(300.2\pm0.5\right)]{K}$. To measure the temperature, it has been written a sketch (code) for Arduino. The sketch allows the user to measure the temperature over time. Then, values of time and temperature are printed on the terminal to be analysed.

The environment temperature has been measured by the room thermometer to ensure a constant temperature during the experiment.

\section{Data analysis and results}\label{Results}
From Eq. \ref{eq:Newtons_cooling}, by defining 

\begin{equation}
y=\ln\left(\frac{T\left(t\right)-T_{env}}{T_{0}-T_{env}}\right)
\label{eq:y_exp}
\end{equation}

we get: 

\begin{equation}
y=kt
\label{eq:linear_function}
\end{equation}

where

\begin{equation}
k=-\frac{1}{\tau}
\label{eq:k_exp}
\end{equation}

\begin{figure}[htbp]
\begin{subfigure}{.5\textwidth}
  \centering
  % include first image
  \includegraphics[height=.60\linewidth]{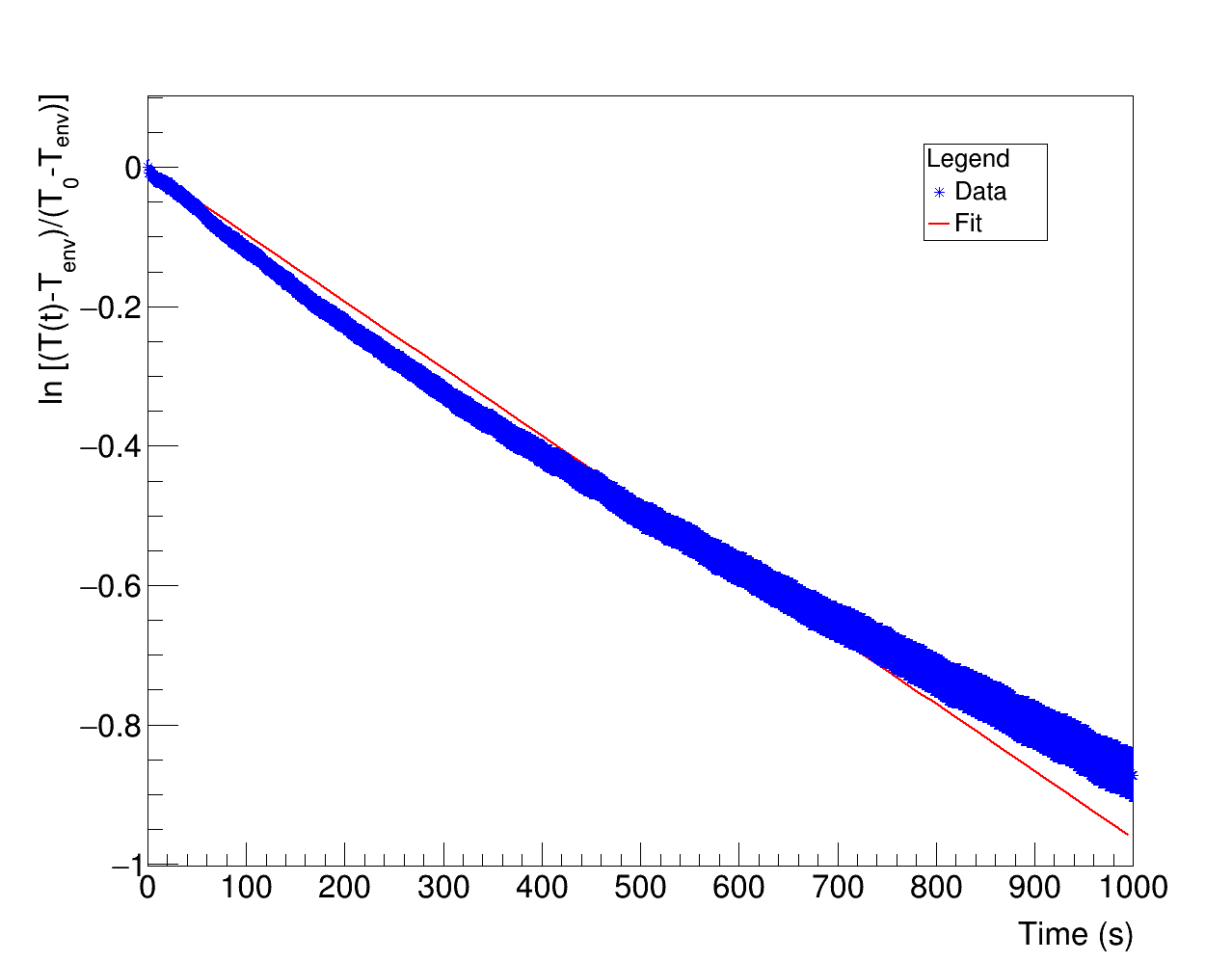}  
  \caption{}
  \label{fig:fit}
\end{subfigure}
\begin{subfigure}{.5\textwidth}
  \centering
  % include second image
  \includegraphics[height=.67\linewidth]{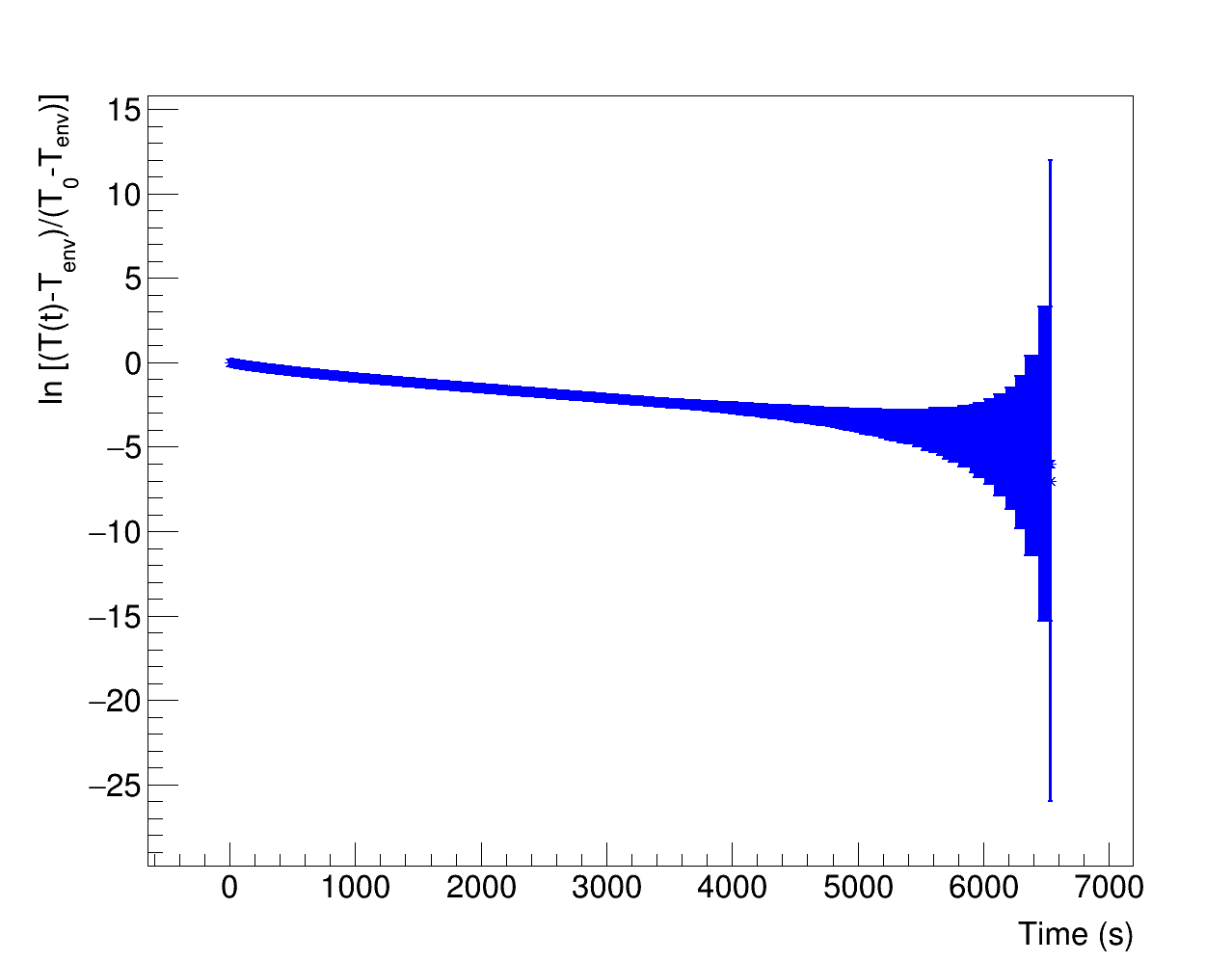}  
  \caption{}
  \label{fig:full}
\end{subfigure}
\caption{(\subref{fig:fit}) Data interpolation. (\subref{fig:full}) Cooling trend up the liquid reaches the room temperature}
\label{fig:CharacteristicCurves}
\end{figure}

It follows that, to estimate the time-constant $k$ by measured data, we can assume the linear dependency (Eq. \ref{eq:linear_function}), interpolate data by it (Fig. \ref{fig:fit}) and estimate the time-constant as: 

\begin{equation}
\tau=-\frac{1}{k}
\label{eq:tau_exp}
\end{equation}

%\begin{figure}[H]
%\centering
%\includegraphics[width=4in]{img/cooling.png}\DeclareGraphicsExtensions.
%\caption{Fit.}
%\label{fig:fit}
%\end{figure}

The slope $k$ resulting by interpolation and the estimated $\tau$ value are given in Tab. \ref{tab:results}.

\begin{table}[hbtp]
\centering
 \caption{ Fit result slope and obtained $\tau$ value.}
 \label{tab:results} 
  \begin{threeparttable}
     \begin{tabular}{ll}
        \toprule
        \(\unit[k]{\left(s^{-1}\right)}\)  & \(\unit[\tau]{\left(s\right)}\) \\
        \midrule
	         $\left(-9.625\pm0.01\right)\cdot10^{-4}$   &  $1039\pm2$\\
        \bottomrule
     \end{tabular}
  \end{threeparttable}
\end{table}

As expected, in Figure \ref{fig:full} we observe that after $5\tau$ the liquid reaches room temperature.

%\begin{figure}[H]
%\centering
%\includegraphics[width=4in]{img/coolingfull.png}\DeclareGraphicsExtensions.
%\caption{Cooling trend up the liquid reaches the room temperature}
%\label{fig:full}
%\end{figure}

\section*{Conclusions}
\addcontentsline{toc}{section}{Conclusions}
In this paper a technique has been presented to measure the Newton’s cooling law time-constant by using an Arduino. The article goal is to encourage teachers to propose the experiment to their students in order to carry on laboratory practice even in this pandemic period. In particular we encourage readers to replicate the experiment changing the body (water in this case) and environment conditions in order to estimate the time-constant for several materials and conditions.

\section*{Acknowledgements}
\addcontentsline{toc}{section}{Acknowledgements}
The author acknowledges the Lab2Go- Fisica \cite{Lab2Gowebpage} collaboration.
\section*{References}
\bibliographystyle{ieeetr}
\bibliography{mybiblio}

\acrodef{RC}[RC]{Resistor Capacitor}

\end{document}